# Georges Lemaître: Life, Science and Legacy. Talk given on 11 November 2026 at the Royal Astronomical Society

*Simon Mitton*

*St Edmund's College, Cambridge CB3 0BN, United Kingdom*

sam11@cam.ac.uk

*www.totalastronomy.com*

*Abstract* This paper celebrates the remarkable life, science and legacy of Abbé Georges Lemaître, the Belgian cleric and professor of physics; he was the architect of the fireworks model for the origin of the universe. He died half a century ago, three days after learning that Arno Penzias and Robert Wilson had discovered the cosmic microwave background. Despite being gravely ill from leukaemia, Lemaître lucidly praised this news, which confirmed the explosive genesis of our universe.

Georges, the first of four sons of Joseph and Marguerite Lemaître, was born on 17 August 1894. He grew up in Belgium's *pays noir*, a dark landscape blighted by the industrial production of glass, coal and steel. His father Joseph founded a glassworks, where we can imagine that from an early age Georges became familiar with fiery furnaces and their smouldering ashes. Georges (10) entered grammar school in Charleroi, where Jesuit teachers thoroughly grounded him in Greek and Latin. By his second year he began to demonstrate outstanding ability in mathematics. After an explosive conflagration reduced the glassworks to sparks and ashes, he family moved to an imposing mid-19[th] century townhouse in Brussels. To complete his high school education, Georges joined 800 boys at Collège St-Michel, an easy 15-minute stroll from home (Lambert 2000).

After one year studying advanced mathematics and science, He began his university career at the Université catholique de Louvain in July 1911 taking the foundation course in engineering, plus a diploma in philosophy. He plunged into analysis and mechanics, and displayed a growing interest in the history of mathematics. He earned distinctions in mathematics and physics, but performed less well in engineering.

On 4 August 1914 the German army invaded neutral Belgium, following the Belgian government's refusal to allow safe passage to France. Lemaître, deeply loyal to "king and country," signed up on 9 August. Six weeks later, the exhausted Belgian army was bottled up along the Yser. Georges fought in the front line, where the infantry held their position from 16 – 31 October, taking casualties of 3,500 killed and 15,000 wounded. The Germans scarpered when Belgian forces flooded the polders During four years of service on the western front. Lemaître kept up his interest in science. He delighted his comrades in arms with accounts of scientific discovery. And he kept up with such developments at an advanced level, mastering Poincaré's *Électricité et Optique*. His bravery won him the Croix de Guerre with Bronze Palms.

On returning to Louvain, dropped civil engineering, to concentrate on mathematics and physics. He gained a doctorate in 1920, at which point he entered a diocesan seminary that fast-tracked men whose vocation had been delayed by military service. During 1921-22 he prepared a personal memoir on the Einstein's physics, which resukted in a major travel award to cover overseas travel 1923–25. Then he won a Fellowship to support study in the United States. In October 1923, after ordination, he went to Cambridge to work with Eddington, lodging at St Edmund's House (now *College*), a residential community of young priests.

When Eddington and Lemaître first met, relativistic cosmology offered two unsatisfactory models, both dating from 1917. Einstein's universe was of finite density, closed, and static; de Sitter's was open, empty, and expanding. Lemaître's period of collaboration resulted in an important paper in which he generalized the definition of simultaneity (Lemaître 1924). Eddington later wrote: "I found M. Lemaître a brilliant student … of great mathematical ability." In August 1924, Lemaître and Eddington participated in the British Association for the Advancement of Science meeting in Toronto, where Eddington gave an acclaimed public lecture on relativity and the bending of starlight.

The next month, 1924 Lemaître arrived at Harvard College Observatory, where he worked on cepheids under Harlow Shapley. In Washington, at a meeting of the American Philosophical Society in April 1925, he presented a paper on the deficiencies of de Sitter's universe. Then, during a meeting at the National Academy, he attended Hubble's famous presentation on the distance of M31, which showed that cepheids provided the key to unlock the extragalactic distance scale. Lemaître arranged to meet Hubble, and learn more about extragalactic distances. Next he went to the Lowell Observatory, where Slipher had measured 36 redshifts.

On returning to teaching duties, Lemaître mounted a major attack on cosmological models (e.g. Lemaître 1925), seeking a solution intermediate between the Einstein and de Sitter universes: a universe containing galaxies (Einstein) and with redshifts (de Sitter). In a stroke of genius, he abandoned static solutions. His paper (Lemaître 1927), on a homogeneous expanding universe of constant mass, appeared in the annals of the scientific society of Brussels, a widely available but seldom read journal . He had rediscovered the solutions by Alexander Friedman (1922). And like Friedman, Lemaitre was given the silent treatment. That is, until 1930, when Eddington leapt into action by publishing a translation in Monthly Notices (Lemaître 1931a). In

Lemaître (1931b) *The beginning of the World from the point of view of quantum theory*, sets out his concept of a primeval atom that shuddered into expansion through radioactive decay. He had his moment of fame in 1931 during the 100th anniversary meeting of the British Association for the Advancement of Science, where he spoke eloquently (lemaître 1931c).

      To summarise Lemaître's legacy, I'll jump to 1948 when Bondi, Gold and Hoyle published their steady-state model, which sparked the Big Bang versus Steady State debates (Mitton 2011). In 1948 Ralph Alpher and Robert Herman had published (Alpher and Herman 1948) a value of 5K for the temperature of the cosmic microwave background (a result for which George Gamow took as much credit as possible). In 1965 the discovery of the CMB was announced: the marked the beginning of the end for steady state model, and the gradual acceptance of explosive (fireworks!) models. The discovery of dark energy rekindled interest in the cosmological constant that Lemaître (1933) had identified with vacuum energy.  In the five decades since his passing, Lemaître's reputation has risen from near obscurity to universal recognition of his achievements as the "Father of the Big Bang."